\documentclass[twocolumn,groupedaddress,superscriptaddress,nofootinbib,longbibliography]{revtex4-1}
\usepackage{graphicx,amsmath,amssymb,amsbsy,txfonts, float}
\usepackage{subfigure,hyperref,bbm,times}
\usepackage[T1]{fontenc}
\usepackage{braket}
\usepackage{epsfig}
\usepackage{color}
\usepackage{graphicx}% Include figure files
\usepackage{dcolumn}% Align table columns on decimal point
\usepackage{bm}% bold math

\providecommand{\openone}{\leavevmode\hbox{\small1\kern-3.8pt\normalsize1}}

\usepackage{soul}

\renewcommand{\tilde}{~}

\newcommand{\id}{1\!\!1}

\newcommand{\kluru}{\ket{L\uparrow,R\uparrow}}
\newcommand{\klurd}{\ket{L\uparrow,R\downarrow}}
\newcommand{\kldru}{\ket{L\downarrow,R\uparrow}}
\newcommand{\kldrd}{\ket{L\downarrow,R\downarrow}}
\newcommand{\bluru}{\bra{L\uparrow,R\uparrow}}

\newcommand{\kaubu}{\ket{A\uparrow,B\uparrow}}
\newcommand{\kaubd}{\ket{A\uparrow,B\downarrow}}
\newcommand{\kadbu}{\ket{A\downarrow,B\uparrow}}
\newcommand{\kadbd}{\ket{A\downarrow,B\downarrow}}
\newcommand{\baubu}{\bra{A\uparrow,B\uparrow}}

\newcommand{\rlr}{\rho_{\text{LR}}}
\newcommand{\rab}{\rho_{\text{AB}}}
\newcommand{\rtilde}{\widetilde{\rho}}

\newcommand{\rd}{\rho_D}
\newcommand{\ku}{\ket{\uparrow}}
\newcommand{\bu}{\bra{\uparrow}}
\newcommand{\kd}{\ket{\downarrow}}
\newcommand{\bd}{\bra{\downarrow}}
\newcommand{\eza}{E_0^A}
\newcommand{\ezb}{E_0^B}
\newcommand{\eoa}{E_1^A}
\newcommand{\eob}{E_1^B}
\newcommand{\eoad}{E_1^{A\,\dagger}}
\newcommand{\eobd}{E_1^{B\,\dagger}}
\newcommand{\kop}{\ket{1_+}}
\newcommand{\kom}{\ket{1_-}}
\newcommand{\ktp}{\ket{2_+}}
\newcommand{\ktm}{\ket{2_-}}
\newcommand{\bop}{\bra{1_+}}
\newcommand{\bom}{\bra{1_-}}
\newcommand{\btp}{\bra{2_+}}
\newcommand{\btm}{\bra{2_-}}
\newcommand{\ab}{\text{AB}}
\newcommand{\dkom}{\ket{\bar{1}_{-}}_\text{D}}
\newcommand{\dkop}{\ket{\bar{1}_{+}}_\text{D}}
\newcommand{\dktm}{\ket{\bar{2}_{-}}_\text{D}}
\newcommand{\dktp}{\ket{\bar{2}_{+}}_\text{D}}

\newcommand{\nkom}{\ket{\bar{1}_{-}}_\text{N}}
\newcommand{\nkop}{\ket{\bar{1}_{+}}_\text{N}}
\newcommand{\nktm}{\ket{\bar{2}_{-}}_\text{N}}
\newcommand{\nktp}{\ket{\bar{2}_{+}}_\text{N}}
\newcommand{\nbom}{\bra{\bar{1}_{-}}_\text{N}}
\newcommand{\nbop}{\bra{\bar{1}_{+}}_\text{N}}
\newcommand{\nbtm}{\bra{\bar{2}_{-}}_\text{N}}
\newcommand{\nbtp}{\bra{\bar{2}_{+}}_\text{N}}
\newcommand{\absp}{\lvert lr'+\eta\,l'r\rvert^2}
\newcommand{\absm}{\lvert lr'-\eta\,l'r\rvert^2}

\hypersetup{
   colorlinks=true,
   linkcolor=blue,
}
\graphicspath{{figure/}}

\begin{document}

\title{Entanglement robustness via spatial deformation of identical particle wave functions}

\author{Matteo Piccolini}
\email{matteo.piccolini@unipa.it}
\affiliation{Dipartimento di Ingegneria, Universit\`{a} di Palermo, Viale delle Scienze, 90128 Palermo, Italy}
\affiliation{INRS-EMT, 1650 Boulevard Lionel-Boulet, Varennes, Qu\'{e}bec J3X 1S2, Canada}

\author{Farzam Nosrati}
\email{farzam.nosrati@unipa.it}
\affiliation{Dipartimento di Ingegneria, Universit\`{a} di Palermo, Viale delle Scienze, 90128 Palermo, Italy}
\affiliation{INRS-EMT, 1650 Boulevard Lionel-Boulet, Varennes, Qu\'{e}bec J3X 1S2, Canada}

\author{Giuseppe Compagno}
\affiliation{Dipartimento di Fisica e Chimica - Emilio Segr\`e, Universit\`a di Palermo, via Archirafi 36, 90123 Palermo, Italy}

\author{Patrizia Livreri}
\affiliation{Dipartimento di Ingegneria, Universit\`{a} di Palermo, Viale delle Scienze, 90128 Palermo, Italy}

\author{Roberto Morandotti}
\affiliation{INRS-EMT, 1650 Boulevard Lionel-Boulet, Varennes, Qu\'{e}bec J3X 1S2, Canada}

\author{Rosario Lo Franco}
\email{rosario.lofranco@unipa.it}
\affiliation{Dipartimento di Ingegneria, Universit\`{a} di Palermo, Viale delle Scienze, 90128 Palermo, Italy}

\begin{abstract}

We address the problem of entanglement protection against surrounding noise by a procedure suitably exploiting spatial indistinguishability of identical subsystems. To this purpose, we take two initially separated and entangled identical qubits interacting with two independent noisy environments. Three typical models of environments are considered: amplitude damping channel, phase damping channel and depolarizing channel. After the interaction, we deform the wave functions of the two qubits to make them spatially overlap before performing spatially localized operations and classical communication (sLOCC) and eventually computing the entanglement of the resulting state. This way, we show that spatial indistinguishability of identical qubits can be utilized within the sLOCC operational framework to partially recover the quantum correlations spoiled by the environment. A general behavior emerges: the higher the spatial indistinguishability achieved via deformation, the larger the amount of recovered entanglement.

\end{abstract}

\maketitle

\section{Introduction}

It is well known that the environment of an open quantum system produces a detrimental noise which has to be dealt with during the implementation of many useful quantum information processing schemes \cite{Preskill2018,Rotter_2015}.
	One of the main goals in the development of fault-tolerant enhanced quantum technologies is to provide a strategy to protect the entanglement from such degradation. This challenge has been addressed, e.g., by the seminal works on quantum error corrections \cite{preskill1998reliable, knill2005quantum, shor1995scheme,steanePRL}, structured environments with memory effects \cite{mazzola2009sudden, bellomo2008entanglementtr,lofrancoreview,aolitareview,XuPRL2010,bylicka2014,lofrancoManSciRep,non-Mar2,non-Mar3,breuerRMP, lofrancoManPRA}, distillation protocols \cite{Bennett1996,kwiatNature,dongNatPhys}, decoherence-free subspaces \cite{zanardi1997noiseless, lidar1998decoherence}, dynamical decoupling and control techniques \cite{Viola1998,viola2005random,darrigo2012AOP, franco2014preserving, orieux2015experimental, Zeno3, lofranco2012PRA, LoFrancoNatCom,damodarakurup2009experimental, cuevas2017cut}.
	
	It is not unusual to find identical particles (i.e., subsystems such as photons, atoms, nuclei, electrons or any artificial qubits of the same species) as building blocks of quantum information processing devices and quantum technologies \cite{obrienreview,AltmanPRXQuantum}. Nonetheless, the standard approach to identical particles based on unphysical labels is known to give rise to formal problems when trying to asses the correlations between constituents with (partially or completely) overlapping spatial wave functions \cite{tichy,ghirardi}. For this reason, many alternative approaches have been developed to deal with the formal aspects of the entanglement of identical particles \cite{ghirardi,facchiIJQI,Li2001PRA,Paskauskas2001PRA,cirac2001PRA,zanardiPRA,eckert2002AnnPhys,balachandranPRL,sasaki2011PRA,giulianoEPJD,bose2002indisting,bose2013,tichyFort,PlenioExtracting,sciaraSchmidt, nolabelappr,compagno2018dealing,slocc,morrisPRX}. Among these, the \emph{no-label} approach \cite{nolabelappr,compagno2018dealing,slocc} provides many advantages: for example, it allows to address the correlations between identical particles exploiting the same tools used for nonidentical ones (e.g., the von Neumann entropy of the reduced density matrix). Furthermore, it provides the known results for distinguishable particles in the limit of non-overlapping (spatially separated) wave functions. Treating the global multiparticle state as a whole, indivisible object, in the no-label approach entanglement strictly depends on both the spatial overlap of the wave functions and on spatially localized measurements. An entropic measure has been recently introduced \cite{indistentanglprotection} to quantify the degree of indistinguishability of identical particles arising from their spatial overlap. Furthermore, an operational framework based on spatially localized operations and classical communication (sLOCC), where the no-label approach finds its natural application, has been firstly theorized \cite{slocc} and later experimentally implemented \cite{experimentalslocc,Barros:20} as a way of activating physical entanglement. Such framework has also been applied to fields such as the exploitation of the Hanbury Brown-Twiss effect with identical particles \cite{Quanta66}, quantum entanglement in one-dimensional systems of anyons \cite{mani2020quantum}, entanglement transfer in a quantum network \cite{castellini2019activating}, and quantum metrology \cite{Castellini2019metrology,Sun_PhaseDiscr}.
	Moreover, in a recent paper \cite{indistentanglprotection} it has been shown that spatial indistinguishability, even partial, can be exploited to recover the entanglement spoiled from the preparation noise of a depolarizing channel. 
	
	In this work, we aim to extend the results of Ref.~\cite{indistentanglprotection} to the wider scenario of different paradigmatic noise channels, namely amplitude damping, phase damping and depolarizing channels, under both Markovian and non-Markovian regimes. To do so, we introduce \emph{spatial deformations}, i.e., transformations turning initially spatially separated (and thus distinguishable) particles into indistinguishable ones by making their wave functions spatially overlap. We then analyze the entanglement dynamics of two identical qubits interacting separately with their own environment, with the goal of showing that the application of the mentioned spatial deformation at a given time of the evolution, immediately followed by the sLOCC measurement, constitutes a procedure capable of recovering quantum correlations. 
	
	This paper is organized as follows: in Section\tilde\ref{general} we introduce the general framework of the analyzed dynamics and the main tools used, namely the deformation operation and the sLOCC protocol. The main results follow in Section\tilde\ref{results}, where we describe the considered model and study the scenarios of an amplitude damping channel, a phase damping channel and a depolarizing channel. Finally, Section~\ref{sec:Discussion} summarizes and discusses the main results.
 
%%%%%%%%%%%%%%%%%%%%%%%%%%%%%%%%%%%%%%%%%%
\begin{figure*}[t!]
		\includegraphics[width=0.7\textwidth]{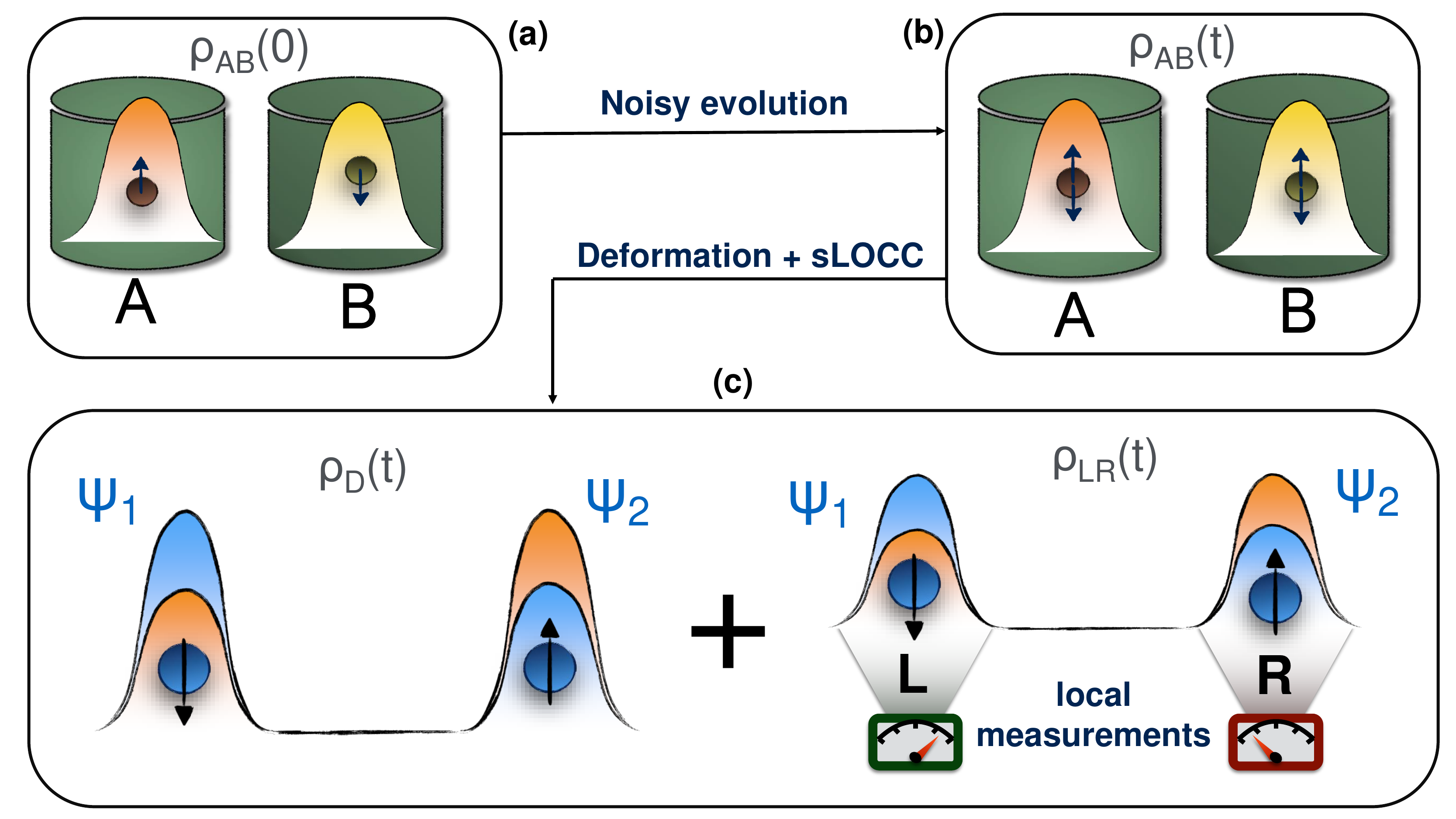}
		\caption{State evolution in the considered scenario. (\textbf{a})\tilde The two qubits are initially prepared in the pure entangled state $\rho_\text{AB}(0)$. (\textbf{b})\tilde They are left to interact with a noisy environment, whose detrimental action produces the mixed state $\rab(t)$. (\textbf{c})\tilde At time $t$ a deformation of the two particles wave functions is performed, immediately followed by a sLOCC measurement.}
		\label{evolution}
	\end{figure*}

\section{Materials and Methods}
\label{general}
	
	In this section we introduce the goal of this paper and the main tools used to achieve it.
	
	Let us consider the following process, illustrated in Fig.\tilde\ref{evolution}: at the beginning, two identical qubits in the entangled state $\rab(0)$ occupy two different regions of space $A$ and $B$, thus being distinguishable and individually addressable. 
Here, they locally interact with two spatially separated and independent noisy environments which spoil the initial correlations. At time $t$, the two particles get decoupled from the environments and undergo a deformation which makes their wave functions spatially overlap into the state $\rd(t)$. Immediately after that, a sLOCC measurement is performed to generate the entangled state $\rlr(t)$. In this work, we show that this procedure allows for the recovery of the entanglement spoiled by the previously introduced noise in an amount which depends on the degree of spatial indistinguishability achieved with the deformation. Three different models of environmental noise shall be considered: an amplitude damping channel, a phase damping channel and a depolarizing channel.
	
	Notice that here the system-environment interaction occurs when the two particles are still distinguishable and no finite time interval separates the deformation from the immediately subsequent sLOCC operation. It will thus be interesting to compare the results of this work with those discussed in Ref.~\cite{indistdynamicalprotection}, where the interaction with the noisy channels happens instead during a finite time interval between the deformation and the sLOCC operation, that is when the qubits are indistinguishable in the frame of the localized environments.
	
	The deformation process bringing two particles to spatially overlap shall be now briefly introduced, followed by a recall of the sLOCC operational framework.
	
	\subsection{Deformations of identical particle states}
	Given a multipartite quantum system, a quantum transformation acting differently on each subpart changing the relations among them is called a \emph{deformation}. In this section we focus on the specific set of continuous deformations which modify the single spatial wave functions of identical particles. In what follows, the no-label formalism \cite{nolabelappr} is used.
	
	Let us take a non-entangled state of two identical particles $\ket{\Phi}=\ket{\phi_1;\phi_2}$, where $\phi_i$ ($i=1,2$) is identified by the values of a complete set of commuting observables describing a spatial wave function $\psi_i$ and an internal degree of freedom $\tau_i$. We suppose that the two particles are initially spatially separated, e.g., localized in two distinct regions $A$ and $B$ such that $\ket{\psi_1^{(0)}}=\ket{A}$, $\ket{\psi_2^{(0)}}=\ket{B}$ and $\braket{A|B}=0$. We want to modify the spatial wave functions of the two particles in order to make them overlap. Thus, we introduce a deformation $\mathcal{D}$ such that
	\begin{equation}
	\label{deformation}
	\ket{\phi_1;\phi_2}
	=\ket{A,\tau_1}\otimes\ket{B,\tau_2}
	\xrightarrow{\mathcal{D}}
	\ket{\psi_1,\tau_1;\psi_2,\tau_2},
	\end{equation}
	with $\braket{\psi_1|\psi_2}\neq 0$.
	Since the two spatially overlapping particles are also identical, they are now \emph{indistinguishable}: their final global state cannot be written as the tensor product of single particle states anymore and must be considered as a whole, i.e. $\ket{\psi_1,\tau_1;\psi_2,\tau_2}\neq\ket{\psi_1,\tau_1}\otimes\ket{\psi_2,\tau_2}$.
	
	A deformation operator acting on identical particles is not, in general, unitary, and its normalized action on a state $\rho$ is thus
	\begin{equation}
	\label{deformationoperator}
	\mathcal{D}[\rho]=\frac{\mathcal{D}\rho\mathcal{D}^\dagger}{\text{Tr}[\mathcal{D}\mathcal{D}^\dagger\rho]}=\sum_i\bar{p}_i\mathcal{D}[\rho_i],
	\end{equation}
	where 
	\begin{equation}
	\bar{p}_i=\frac{\text{Tr}[\mathcal{D}\mathcal{D}^\dagger\rho_i]}{\text{Tr}[\mathcal{D}\mathcal{D}^\dagger\rho]},
	\:\:\: \mathcal{D}[\rho_i]=\frac{\mathcal{D}\rho_i\mathcal{D}^\dagger}{\text{Tr}[\mathcal{D}\mathcal{D}^\dagger\rho_i]}.
	\end{equation}

	\subsection{sLOCC, Spatial Indistinguishability and Concurrence}
	The natural extension of the standard local operation and classical communication framework (LOCC) for distinguishable particles to the scenario of indistinguishable (and thus individually unaddressable) particles is provided by the \emph{spatially localized operations and classical communication} (sLOCC)  environment \cite{slocc}. Given a set of indistinguishable particles, sLOCC consist in a projective measurement of the global state over distinct spatially separated regions, followed by a post-selection of the outcomes where only one particle is found in each location.
	The result of this operation is an entangled state whose physical accessibility has been demonstrated in a quantum teleportation experiment \cite{experimentalslocc}.
	
	Suppose we are given a state $\rho$ of two identical and indistinguishable particles, e.g., obtained by the application\tilde\eqref{deformationoperator} of the deformation\tilde\eqref{deformation}, and assume they have pseudo-spin $1/2$.
	The whole sLOCC operation (projection and post-selection) amounts to projecting the two qubits state on the subspace spanned by the basis
	\begin{equation}
	\label{basis}
	\mathcal{B}_{LR}=\{\kluru,\klurd,\kldru,\kldrd\},
	\end{equation}
	via the projection operator
	\begin{equation}
	\label{sloccprojector}
	\hat{\Pi}_{LR}
	=\sum_{\sigma,\tau=\uparrow,\downarrow}\ket{L\sigma,R\tau}\bra{L\sigma,R\tau}.
	\end{equation}
	Since the constituents are indistinguishable before the detection, it is impossible to know exactly which particle will be found in which region. The sLOCC operation generates the (normalized) two-particle entangled state
	\begin{equation}
	\label{sloccstate}
	\rlr(t)
	=\frac{\hat{\Pi}_{LR}\,\rho(t)\,\hat{\Pi}_{LR}}{\text{Tr}\left[\hat{\Pi}_{LR}\,\rho\right]},
	\end{equation}
	with probability
	\begin{equation}
	\label{sloccprob}
	P_\text{LR}
	=\text{Tr}\left[\hat{\Pi}_{LR}\,\rho\right].
	\end{equation}
	After the sLOCC measurement, the two qubits occupy two distinct regions of space and are thus now distinguishable and individually addressable. Furthermore, since in the no-label formalism the inner product between two-particle states is given by the rule \cite{nolabelappr}
	\begin{equation}
	\braket{\phi'_1;\phi'_2|\phi_1;\phi_2}
	=\braket{\phi'_1|\phi_1}\braket{\phi'_2|\phi_2}+\eta\braket{\phi'_1|\phi_2}\braket{\phi'_2|\phi_1},
	\end{equation}
	with $\eta=1$ for bosons and $\eta=-1$ for fermions, particle statistics naturally emerges within the sLOCC framework and is thus expected to play a role in the dynamics.
	
	The sLOCC scenario also allows for the introduction of an entropic measure of the particles' indistinguishability after the deformation\tilde\eqref{deformation}, which depends on the achieved spatial distribution of their wave functions $\psi_1,\,\psi_2$ over the two regions $L$ and $R$ where sLOCC measurement occurs. Given the probability $P_{X\psi_i}$ of finding the qubit having wave function $\psi_i$ ($i=1,2$) in the region $X$ ($X=L,R$), the spatial indistinguishability measure is given by \cite{indistentanglprotection}
	\begin{equation}
	\label{indistinguishability}
	\mathcal{I}=
	-\dfrac{P_{L\psi_1}P_{R\psi_2}}{\mathcal{Z}} \log_2 \dfrac{P_{L\psi_1}P_{R\psi_2}}{\mathcal{Z}}
	-\dfrac{P_{{L}\psi_2}P_{{R}\psi_1}}{\mathcal{Z}}\log_2 \dfrac{P_{{L}\psi_2}P_{{R}\psi_1}}{\mathcal{Z}},
	\end{equation}
	where $\mathcal{Z}=P_{{L\psi_1}}P_{{R}\psi_2}+P_{{L}\psi_2}P_{{R}\psi_1}$.
	Notice that\tilde\eqref{indistinguishability} ranges from $0$ for spatially separated (thus distinguishable) particles (e.g. when $P_{L\psi_1}=P_{R\psi_2}=1$) to $1$ for maximally indistinguishable particles ($P_{L\psi_1}=P_{L\psi_2},\,P_{R\psi_1}=P_{R\psi_2}$). Hereafter, we assume for convenience that the spatial wave functions of the single indistinguishable particles after the deformation have the form
	\begin{equation}
	\label{wfstructure}
	\ket{\psi_1}=l\ket{L}+r\ket{R},\quad
	\ket{\psi_2}=l'\ket{L}+r'\ket{R},
	\end{equation}
	where
	\begin{equation}
	\label{slocccoeff}
	l=\braket{L|\psi_1},\,r=\braket{R|\psi_1},\,l'=\braket{L|\psi_2},\,r'=\braket{R|\psi_2}
	\end{equation}
	are complex coefficients such that $|l|^2+|r|^2=|l'|^2+|r'|^2=1$. In the following analysis, we shall conveniently set $l=r'$ to assure that the sLOCC probability $P_\text{LR}$ is different from zero.
	
	As previously stated, the state $\rlr$ obtained by the sLOCC measurement is entangled. We recall that the entanglement of the bipartite quantum state $\rlr$ of two distinguishable qubits can be quantified by the Wootters concurrence \cite{concurrence,indistentanglprotection}
	\begin{equation}
	\label{concurrence}
	C(\rlr)=\max \{0,\sqrt{\lambda_4}-\sqrt{\lambda_3}-\sqrt{\lambda_2}-\sqrt{\lambda_1}\},
	\end{equation}
	where $\lambda_i$ are the eigenvalues in decreasing order of the matrix $\xi=\rlr\,\rtilde_{\text{LR}}$, with $\rtilde_{\text{LR}}=(\sigma^{\text{L}}_{y}\otimes\sigma_{y}^{\text{R}})\,\rab^{*}\,(\sigma^{\text{L}}_{y}\otimes\sigma_{y}^{\text{R}})$ and $\sigma_{y}^{\text{L}}$, $\sigma_{y}^{\text{R}}$ being the usual Pauli matrix $\sigma_y$ localized, respectively, on the particle in $\text{L}$ and in $\text{R}$.

	\section{Indistinguishability as a feature for recovering entanglement} \label{results}
	
	In this section we report our main results.
Each of the two independent environments is modeled as a bath of harmonic oscillators in the vacuum state except for one mode which is coupled to the qubit interacting with it. 
%This is indeed the Jaynes-Cummings model describing, e.g., a two-level atom on resonance with a single mode of a quantized electromagnetic field in a cavity supporting the transition frequency $\omega_0$ of the qubit. 
Considering a qubit-cavity model with just one excitation overall allows us to treat the reservoir as characterized by a Lorentzian spectral density \cite{breuer2002theory,Haikka_2010}
	\begin{equation}
	\label{lorentzdensity}
	J(\omega)=
	\frac{\gamma}{2\pi}
	\frac{\lambda^2}{(\omega-\omega_0)^2+\lambda^2},
	\end{equation}
	where $\omega_0$ is the qubit transition frequency, $\gamma$ is the microscopic system-environment coupling constant related to the decay of the excited state of the qubit in the Markovian limit of flat spectrum, and $\lambda$ is the spectral width of the coupling quantifying the leakage of photons through the cavity walls. 
	The relaxation time $\tau_R$ on which the state of the system changes is related to the coupling constant by the relation $\tau_R\approx\gamma^{-1}$, while the reservoir correlation time $\tau_B$ is connected to the spectral width of the coupling by $\tau_B\approx\lambda^{-1}$. These coefficients regulate the behavior of the system: when $\gamma<\lambda/2\,(\tau_R>2\tau_B)$ the system is weakly coupled to the environment, the reservoir correlation time is shorter than the relaxation time and we are in a Markovian regime; when $\gamma>\lambda/2\,(\tau_R<2\tau_B)$ instead, we are in the strong coupling scenario, where the relaxation time is shorter than the bath correlation time and the regime is non-Markovian. The way each qubit interacts with its own reservoir depends on the type of noise channel taken into account.  
	
	The action of the three noisy channels considered in this paper shall be computed within the usual Kraus operators formalism, or operator-sum representation \cite{nielsen2010quantum}. The general expression of the single-qubit evolved density matrix is then given by $\rho(t)=\sum_i E_i \rho(0) E_i^\dagger$, where the $E_i$'s are the time-dependent Kraus operators corresponding to the specific channel and depend on the disturbance probability (decoherence function) $p(t)$.
	Each channel in fact introduces a time-dependent disturbance on the system with a probability $p(t)=1-q(t)$ obtained by solving the differential equation \cite{breuer2002theory,Bellomo_2007}
	\begin{equation}
	\label{probability}
	\dot{q}(t)
	=-\int_{0}^{t}dt_1\,f(t-t_1)\,q(t_1),
	\end{equation}
	where the correlation function $f(t-t_1)$ is given by the Fourier transform of the spectral density $J(\omega)$ of the reservoir, namely
	\begin{equation}
	f(t-t_1)=
	\int d\omega\,J(\omega)\,e^{-i(\omega-\omega_0)(t-t_1)}.
	\end{equation}
	Solving Eq.\tilde\eqref{probability} for the spectral density\tilde\eqref{lorentzdensity}, one obtains the disturbance (or error) probability \cite{breuer2002theory}
	\begin{equation}
	\label{probabilityfunc}
	p(t)
	=1-e^{-\lambda t}\left[\cos\left(\frac{d\,t}{2}\right)+\frac{\lambda}{d}\sin\left(\frac{d\,t}{2}\right)\right]^2,
	\end{equation}
	with $d=\sqrt{2\gamma\lambda-\lambda^2}$.
	Notice that this solution encompasses both Markovian and non-Markovian regimes, depending on the ratio $\lambda/\gamma$. In particular, in the Markovian limit of flat spectrum which occurs for $\gamma/\lambda\ll 1$, it is straightforward to see that $p(t)=1-e^{-\gamma t/2} $, as expected  \cite{nielsen2010quantum}.
	%The solution for the Markovian regime can be obtained by simply changing the trigonometric functions with the corresponding hyperbolic ones and by switching $d$ with $i\,d$.
	In general, the error probability\tilde\eqref{probabilityfunc} is such that $p(0)=0$ and $\underset{t\to\infty}{\lim}\,p(t)=1$.

	\subsection{Amplitude Damping Channel}
	
	The action of the amplitude damping channel on a single qubit in the operator-sum representation is given by the Kraus operators \cite{nielsen2010quantum}
	\begin{equation}
	\label{krausadc}
	\begin{gathered}
	E_0=\ku\bu+\sqrt{1-p(t)}\,\kd\bd
	=E_0^{\dagger},
	\\
	E_1=\sqrt{p(t)}\,\ku\bd,
	\quad
	E_1^{\dagger}=\sqrt{p(t)}\,\kd\bu.
	\end{gathered}
	\end{equation}
	Consider two identical qubits initially prepared in the Bell singlet state
	\begin{equation}
	\label{bellsinglet}
	\kom_{\text{AB}}
	=\frac{1}{\sqrt{2}}\Big(\kaubd-\kadbu\Big),
	\end{equation}
	with $A$ and $B$ being two distinct spatial regions ($\braket{A|B}=0$). Thanks to the fact that the the two environmental interactions are independent, the state after the noisy interaction is given by
	\begin{equation}
	\label{noisyevolution}
	\begin{aligned}
	\rab(t)
	=&\left(\eza\otimes\ezb\right)\rab(0)\left(\eza\otimes\ezb\right)\\
	+&\left(\eoa\otimes\eob\right)\rab(0)\left(\eoad\otimes\eobd\right)\\
	+&\left(\eza\otimes\eob\right)\rab(0)\left(\eza\otimes\eobd\right)\\
	+&\left(\eoa\otimes\ezb\right)\rab(0)\left(\eoad\otimes\ezb\right),
	\end{aligned}
	\end{equation}
	where $E_i^X$ ($i=1,2,\,X=A,B$) denotes the $i$-th single particle Kraus operator of Eq.\tilde\eqref{krausadc} acting on the qubit localized in region $X$, while 
	$\rab(0)=\kom_{\text{AB}}\bom_{\text{AB}}$ is the initial density matrix.
%	\begin{equation}
%	\label{initialstate}
%	\begin{aligned}
%	\rab(0)&
%	=\kom_{\text{AB}}\bom_{\text{AB}}\\&
%	=\frac{1}{2}\Big(\kaubd\baubd+\kadbu\badbu\\&
%	\quad\,\,\,
%	-\kaubd\badbu-\kadbu\baubd\Big)
%	\end{aligned}
%	\end{equation}
	Using Eq.\tilde\eqref{krausadc} in the above equation, one then finds 
	\begin{equation}
	\label{noisystate}
	\rab(t)
	=\Big(1-p(t)\Big)\kom_\ab\bom_\ab
	+p(t)\kaubu\baubu.
	\end{equation}
	We now want to apply the deformation defined in Eq.\tilde\eqref{deformation} to the state \tilde\eqref{noisystate} at time $t$. State $\kom_\ab$ gets mapped to
	\begin{equation}
	\label{deformed1-}
	\dkom=\frac{1}{\sqrt{2}}\Big(\ket{\psi_1\uparrow,\psi_2\downarrow}-\ket{\psi_1\downarrow,\psi_2\uparrow}\Big),
	\end{equation}
	which is not a normalized state since $\braket{\psi_1|\psi_2}\neq0$. In order to write it in terms of a normalized state $\nkom$, we compute
	\begin{equation}
	\label{coeff1}
	\braket{\bar{1}_-|\bar{1}_-}_\text{D}
	=C_1^2,
	\quad
	C_1:=\sqrt{1-\eta\lvert\braket{\psi_1|\psi_2}\rvert^2},
	\end{equation}
	and write it as
	\begin{equation}
	\dkom=C_1\nkom.
	\end{equation}
	The same is done for the deformation of $\kaubu$, which gets mapped to 
	\begin{equation}
	\label{coeff2}
	\ket{\psi_1\uparrow,\psi_2\uparrow}_\text{D}
	=C_2\ket{\psi_1\uparrow,\psi_2\uparrow}_\text{N},
	\quad
	C_2:=\sqrt{1+\eta\lvert\braket{\psi_1|\psi_2}\rvert^2},
	\end{equation}
	where
	\begin{equation}
	\braket{\psi_1\uparrow,\psi_2\uparrow|\psi_1\uparrow,\psi_2\uparrow}_\text{N}=1.
	\end{equation}
	The normalized state resulting from the spatial deformation\tilde\eqref{deformationoperator} of the state \tilde\eqref{noisystate} is thus
\begin{widetext}	
	\begin{equation}
	\label{deformedstate}
	\rho_\text{D}(t)
	=\frac{\Big(1-p(t)\Big)C_1^2\nkom\nbom
		+p(t)\,C_2^2\ket{\psi_1\uparrow,\psi_2\uparrow}_\text{N}\bra{\psi_1\uparrow,\psi_2\uparrow}_\text{N}}
	{\Big(1-p(t)\Big)C_1^2+p(t)\,C_2^2}.
	\end{equation}
	\end{widetext}
	Following the scheme shown in Figure\tilde\ref{evolution}, we perform the sLOCC measurement immediately after the deformation, applying the projection operator\tilde\eqref{sloccprojector} onto the state\tilde\eqref{deformedstate}, which finally gives 
	\begin{widetext}	
	\begin{equation}
	\label{rhoslocc}
	\rlr(t)
	=\frac{\Big(1-p(t)\Big)\absm\kom_\text{LR}\bom_\text{LR}
	+p(t)\,\absp\kluru\bluru}
	{\Big(1-p(t)\Big)\absm+p(t)\,\absp},
	\end{equation}
	\end{widetext}	
	where $l,r,l',r'$ are the wave function coefficients defined in\tilde\eqref{slocccoeff}.
	
	\begin{figure}[b!]
		\includegraphics[width=0.48\textwidth]{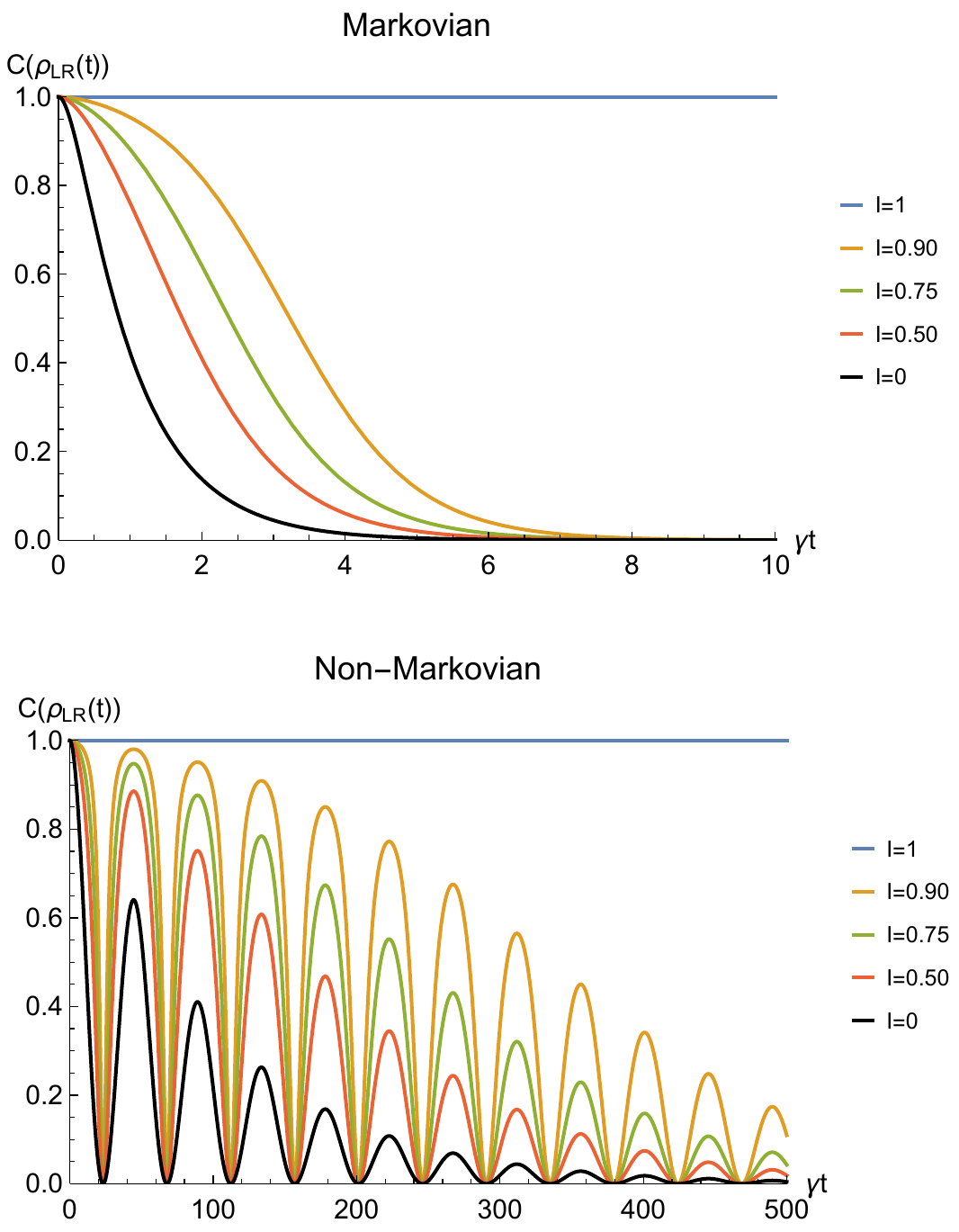}
		\caption{Concurrence of two identical qubits (fermions with $l,l',r,r'>0$, bosons with one of these four coefficients negative) in the initial state $\kom_\text{AB}$ subjected to localized amplitude damping channels, undergoing an instantaneous deformation+sLOCC operation at time $t$ for different degrees of spatial indistinguishability $\mathcal{I}$ (with $|l|=|r'|$). Both the Markovian ($\lambda=5\gamma$) (upper panel) and non-Markovian ($\lambda=0.01\gamma$) (lower panel) regimes are reported.}
		\label{amplitude_damping}
	\end{figure}
	
	\begin{figure}[b!]
		\includegraphics[width=0.48\textwidth]{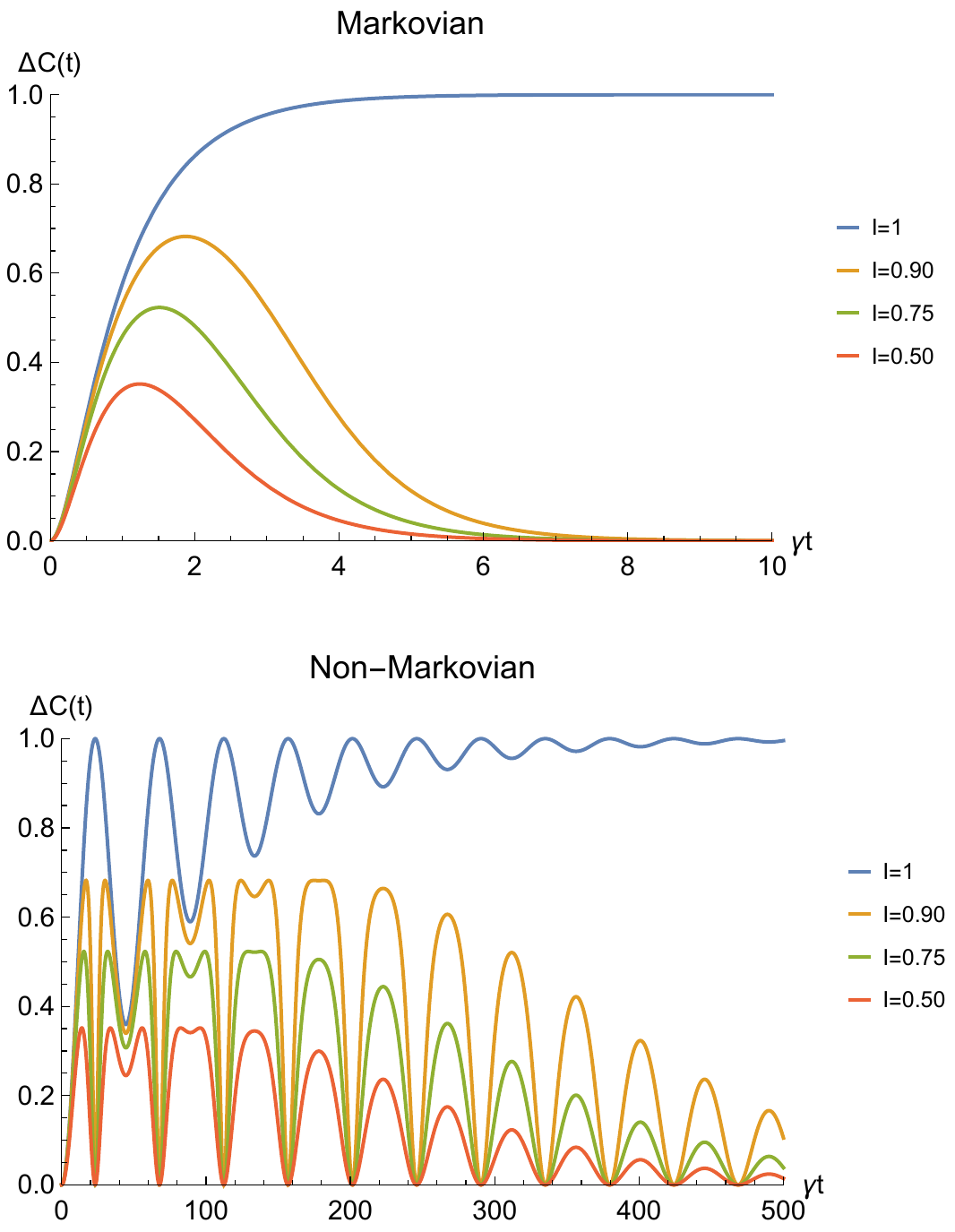}
		\caption{Net gain in the entanglement recovery of two identical qubits (fermions with $l,l',r,r'>0$, bosons with one of these four coefficients negative) in the initial state $\kom_\text{AB}$ under localized amplitude damping channels, thanks to the deformation+sLOCC operation performed at time $t$. Results are reported for different degrees of spatial indistinguishability $\mathcal{I}$ (with $|l|=|r'|$). Both the Markovian ($\lambda=5\gamma$) (upper panel) and non-Markovian ($\lambda=0.01\gamma$) (lower panel) regimes are shown.}
		\label{amplitude_damping_diff}
	\end{figure}
	
	In order to study the entanglement evolution of the state $\rlr(t)$ of Eq.\tilde\eqref{rhoslocc}, we calculate the concurrence defined in Eq.\tilde\eqref{concurrence}, which is
	\begin{equation}
	\label{concurrencead}
	C\big(\rlr(t)\big)
	=\frac{\absm\Big(1-p(t)\Big)}
	{\absm\Big(1-p(t)\Big)+\absp\,p(t)},
	\end{equation}
	where the statistics parameter $\eta$ explicitly appears, as expected. 
	As a first consideration, we notice that the results about entanglement dynamics for bosons can be obtained from the ones for fermions (and vice versa) by simply changing sign to one of the coefficients $l,\,r,\,l',\,r'$ (that is, by shifting the phase of one of them by $\pi$). Therefore, in order to fix a framework to analyze the concurrence, we assume we are dealing with fermions whose spatial wave functions are distributed over the regions $L$ and $R$ with positive real coefficients. This reasoning shall hold for the other noisy channels, so that the presented results are also valid for bosons. With this assumption, we get the concurrence as
	\begin{equation}
	\label{concurrenceadreal}
	C\big(\rlr(t)\big)
	=\frac{\Big[(lr')^2+(l'r)^2+2\,ll'rr'\Big]\Big(1-p(t)\Big)}
	{(lr')^2+(l'r)^2+2\,ll'rr'\Big(1-2p(t)\Big)}.
	\end{equation}
	We point out that when no deformation is performed and the particles remain distinguishable in two distinct regions ($\mathcal{I}=0$), the sLOCC projector\tilde\eqref{sloccprojector} is equivalent to the identity operator. This implies that when the particles are not brought to spatially overlap, our procedure gives the same entanglement we would have without performing the sLOCC operation. For this reason, we take the results for $\mathcal{I}=0$ as the term of comparison to quantify the entanglement gained due to the \emph{deformation + sLOCC procedure}, i.e., $\Delta C(t):=C\big(\rlr(t)\big)-C\big(\rab(t)\big)$.
	Figure\tilde\ref{amplitude_damping} shows the concurrence\tilde\eqref{concurrenceadreal} for both the Markovian and the non-Markovian regimes, while Figure\tilde\ref{amplitude_damping_diff} displays $\Delta C(t)$.
	
	As can be seen in Figure\tilde\ref{amplitude_damping}, spatial indistinguishability\tilde\eqref{indistinguishability} has a direct influence on the general behavior: when the particles are not perfectly indistinguishable ($\mathcal{I}\neq1$), the entanglement vanishes with a monotonic decay in the Markovian regime and with a periodic one in the non-Markovian regime.
	From Figure\tilde\ref{amplitude_damping_diff}, we can see that when $\mathcal{I}\neq1$ the deformation and sLOCC procedure becomes inefficient in recovering the correlations as time grows. Nonetheless, it is interesting to notice that it provides an initial effective advantage as a consequence of the fact that the decay rate shown in Figure\tilde\ref{amplitude_damping} gets lower as the indistinguishability increases.
	However, when the particles wave functions maximally overlap ($\mathcal{I}=1$), the entanglement remains stable at its initial maximum value, thus becoming unaffected by the noise.
	These results show that, in the scenario of the amplitude damping channel, we have provided an operational framework where spatial indistinguishability, even imperfect, of two identical qubits can be exploited as a scheme to recover quantum correlations spoiled by a short-time interaction with the noisy environment.
	
	\begin{figure}[t!]
		\includegraphics[width=0.48\textwidth]{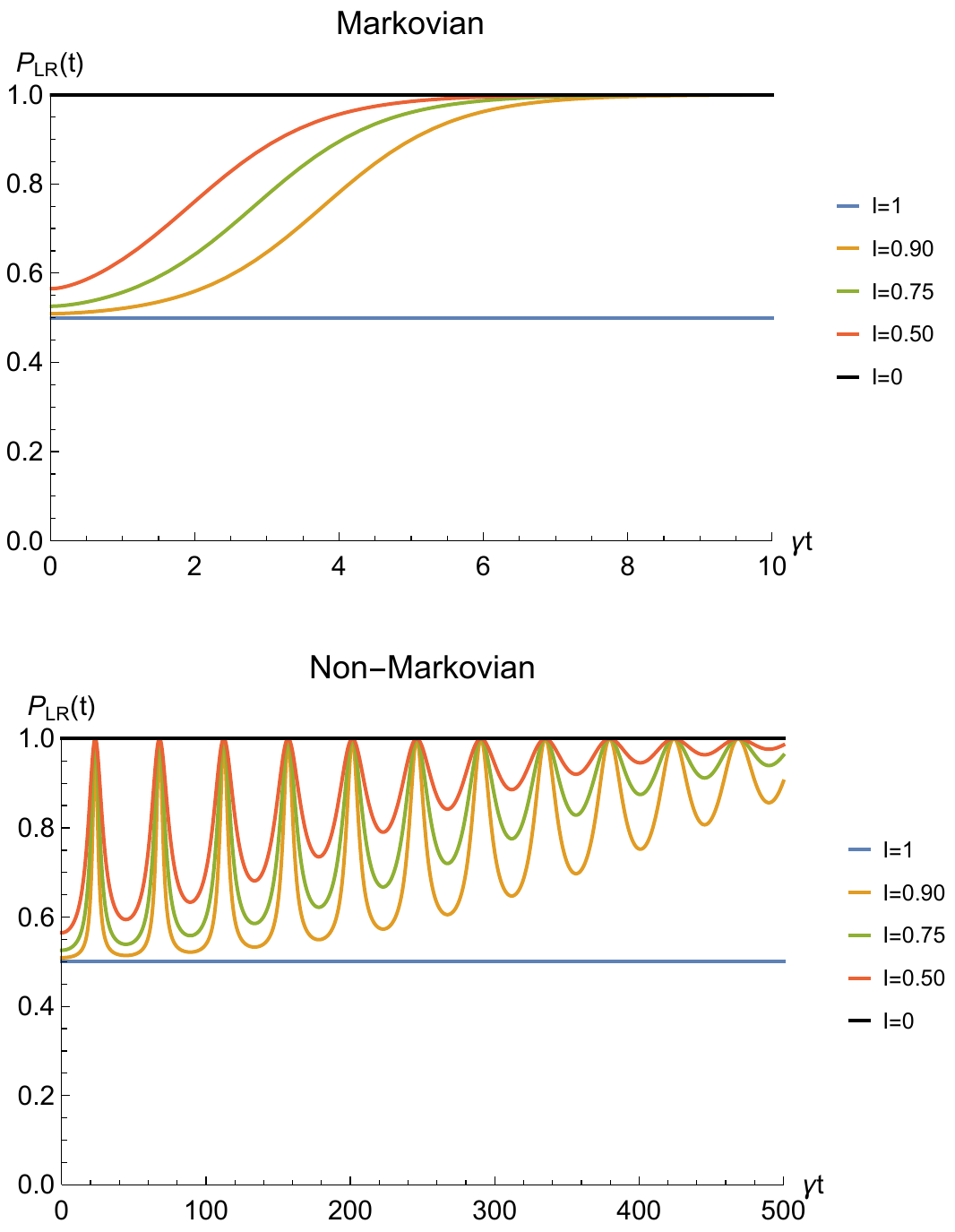}
		\caption{Success probability of obtaining a nonzero outcome from the sLOCC projection for fermions ($l,l',r,r'>0$ and $l=r'$) interacting with localized amplitude damping channels. Different degrees of spatial indistinguishability are reported in both the Markovian ($\lambda=5\gamma$) (upper panel) and non-Markovian ($\lambda=0.01\gamma$) (lower panel) regimes.}
		\label{sloccprobad}
	\end{figure}
	
	Finally, to check whether such procedure would be of any practical interest we have to analyze its theoretical probability of success. This strictly depends on the probability for the sLOCC projection\tilde\eqref{sloccstate} to produce a non-null result, physically representing a state which does not get discarded during the postselection. Such probability is defined in Eq.\tilde\eqref{sloccprob} and, for identical qubits undergoing a local interaction with an amplitude damping channel, it is equal to
\begin{equation}
	\label{eqsloccprobad}
	P_\text{LR}(t)
	=\frac{(lr')^2+(l'r)^2-2\eta\,ll'rr'\Big(1-2\,p(t)\Big)}
	{C_1^2\Big(1-p(t)\Big)+C_2^2\,p(t)}.
	\end{equation}
	Figure\tilde\ref{sloccprobad} shows the success probability\tilde\eqref{eqsloccprobad} for different degrees of spatial indistinguishability in both the Markovian and non-Markovian regime in the case of fermions. As can be seen, when the indistinguishability is not maximum, the probability of success tends to $1$ as time passes in both regimes, thus giving rise to a trade-off with the concurrence. The trade-off is confirmed by the probability being constant and equal to $1/2$ when the concurrence is maximum, i.e. for $\mathcal{I}=1$. For bosons, the time-dependent success probability corresponding to $\mathcal{I}=1$ (with the constraint $l=r'=l'=-r$) and to the concurrence plotted in Fig.~\ref{amplitude_damping} is $P_\text{LR}(t)=1-p(t)$ (notice, however, that this success probability can be improved by differently setting the coefficients of the spatial wave functions).

	\subsection{Phase Damping channel}
	
		\begin{figure}[t!]
		\includegraphics[width=0.48\textwidth]{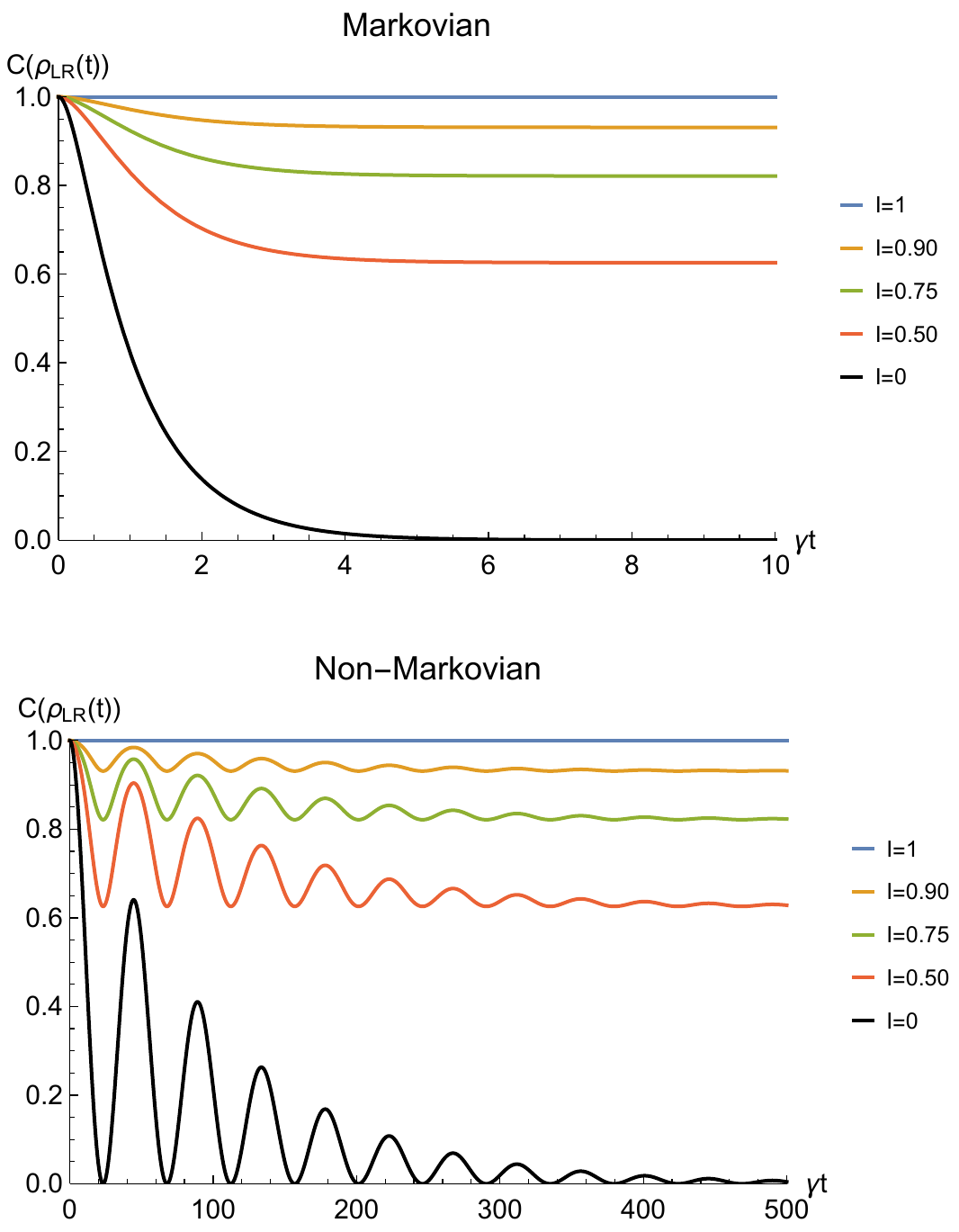}
		\caption{Concurrence of two identical qubits (fermions with $l,l',r,r'>0$, bosons with one of these four coefficients negative) in the initial state $\kom_\text{AB}$ interacting with localized phase damping channels, undergoing an instantaneous deformation+sLOCC operation at time $t$ for different degrees of spatial indistinguishability $\mathcal{I}$ (with $|l|=|r'|$). Both the Markovian ($\lambda=5\gamma$) (upper panel) and non-Markovian ($\lambda=0.01\gamma$) (lower panel) regimes are reported.}
		\label{phase_damping}
	\end{figure}
	
	A phase damping channel acting on a single qubit is described by the Kraus operators
	\begin{equation}
	\label{krauspdc}
	\begin{gathered}
	E_0
	=\ket{\uparrow}\braket{\uparrow}+\sqrt{1-p(t)}\ket{\downarrow}\bra{\downarrow}=E_0^\dagger,\\
	E_1=\sqrt{p(t)}\ket{\downarrow}\bra{\downarrow}=E_1^\dagger.
	\end{gathered}
	\end{equation}
	Once again, we consider the Bell state $\kom_\text{AB}$ of two identical qubits defined in\tilde\eqref{bellsinglet} as our initial state. The evolved state $\rab(t)$ after the interaction with the two independent environments is computed as in Eq.\tilde\eqref{noisyevolution}, which for the phase damping channel described by the above Kraus operators gives
	\begin{equation}
	\label{noisystate2}
	\rab(t)
	=\left(1-\frac{p(t)}{2}\right)\kom_\text{AB}\bom_\text{AB}
	+\frac{p(t)}{2}\kop_\text{AB}\bop_\text{AB},
	\end{equation}
	where $\kop_\text{AB}$ is the Bell state defined as
	\begin{equation}
	\label{belltriplet}
	\kop_{\text{AB}}
	=\frac{1}{\sqrt{2}}\Big(\kaubd+\kadbu\Big).
	\end{equation}
	At time $t$, deformation\tilde\eqref{deformation} is applied to the state\tilde\eqref{noisystate2} to make the two particles spatially overlap. Deformation of $\kom_\text{AB}$ gives the state\tilde\eqref{deformed1-}, while $\kop_\text{AB}$ gets mapped to
	\begin{equation}
	\label{deformed1+}
	\dkop=\frac{1}{\sqrt{2}}\Big(\ket{\psi_1\uparrow,\psi_2\downarrow}+\ket{\psi_1\downarrow,\psi_2\uparrow}\Big).
	\end{equation}
	Once again, state $\dkop$ is not normalized: it is indeed easy to show that
	\begin{equation}
	\dkop=C_2\nkop,
	\end{equation}
	where $\braket{\bar{1}_+|\bar{1}_+}_N=1$ and $C_2$ is defined in\tilde\eqref{coeff2}. Thus, the global normalized state after the deformation is
	\begin{equation}
	\label{deformedstate2}
	\rho_D(t)
	=\frac{\Big(1-\frac{1}{2}\,p(t)\Big)C_1^2\,\nkom\nbom
		+\frac{1}{2}\,p(t)\,C_2^2\,\nkop\nbop}
	{\Big(1-\frac{1}{2}\,p(t)\Big)C_1^2
		+\frac{1}{2}\,p(t)\,C_2^2}.
	\end{equation}
	Finally, the sLOCC operation is performed: the action of the projection operator\tilde\eqref{sloccprojector} on the state\tilde\eqref{deformedstate2}, as defined in Eq.\tilde\eqref{sloccstate}, gives
	\begin{widetext}	
	\begin{equation}
	\label{rhoslocc2}
	\begin{aligned}
	\rlr(t)&
	=\frac{\left(1-\frac{1}{2}\,p(t)\right)\absm
	\kom_\text{LR}\bom_\text{LR}
	+\frac{1}{2}\,p(t)\absp\kop_\text{LR}\bop_\text{LR}}{\left(1-\frac{1}{2}\,p(t)\right)\absm
	+\frac{1}{2}\,p(t)\absp}.
	\end{aligned}
	\end{equation}
	\end{widetext}	
	
	We now study the entanglement evolution of such a state by the concurrence $C(\rlr(t))$, which is readily found to be
	\begin{equation}
	\label{concurrencepd}
	\begin{gathered}
	C\big(\rlr(t)\big)
	=\max\left\{0,\,\lambda_1(t)-\lambda_2(t)\right\},\\
	\lambda_1(t):
	=\max\left\{\lambda_A(t),\,\lambda_B(t)\right\},
	\lambda_2(t):
	=\min\left\{\lambda_A(t),\quad
	\lambda_B(t)\right\},
	\end{gathered}
	\end{equation}
	with
	\[
	\lambda_A(t):
	=\frac{\left(1-\frac{1}{2}\,p(t)\right)\absm}
	{\left(1-\frac{1}{2}\,p(t)\right)\absm
		+\frac{1}{2}\,p(t)\absp},
	\]
	\[
	\lambda_B(t):
	=\frac{\frac{1}{2}\,p(t)\,\absp}
	{\left(1-\frac{1}{2}\,p(t)\right)\absm
		+\frac{1}{2}\,p(t)\absp}.
	\]
	Focusing the analysis once again on fermions with real and positive coefficients $l,r,l',r'$ to fix a framework, concurrence\tilde\eqref{concurrencepd} is then equal to
	\begin{equation}
	\label{concurrencepdreal}
	C\big(\rlr(t)\big)
	=\frac{\Big(1-p(t)\Big)\Big[(lr')^2+(l'r)^2\Big]
		+2ll'rr'}
	{(lr')^2+(l'r)^2+\Big(1-p(t)\Big)2ll'rr'}.
	\end{equation}
	
	\begin{figure}[t]
		\includegraphics[width=0.48\textwidth]{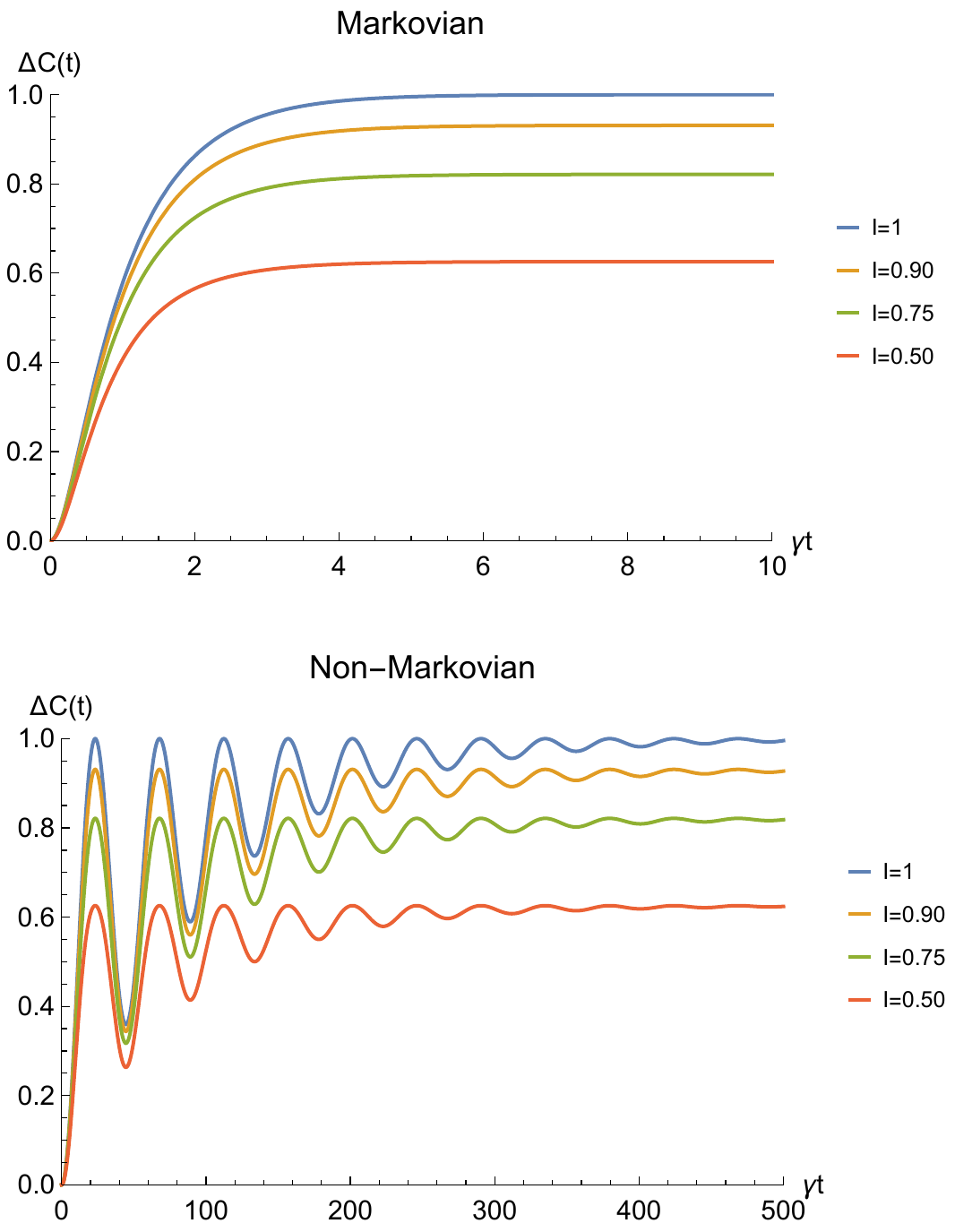}
		\caption{Net gain in the entanglement recovery of two identical qubits (fermions with $l,l',r,r'>0$, bosons with one of these four coefficients negative) in the initial state $\kom_\text{AB}$ under localized phase damping channels, thanks to the deformation+sLOCC operation performed at time $t$. Results are reported for different degrees of spatial indistinguishability $\mathcal{I}$ (with $|l|=|r'|$). Both the Markovian ($\lambda=5\gamma$) (upper panel) and non-Markovian ($\lambda=0.01\gamma$) (lower panel) regimes are shown.}
		\label{phase_damping_diff}
	\end{figure}

The time behavior of the concurrence of Eq.\tilde\eqref{concurrencepdreal} is plotted in Figure\tilde\ref{phase_damping} for both the Markovian and the non-Markovian regime, while the net gain due to the deformation and sLOCC operation is depicted in Figure\tilde\ref{phase_damping_diff}.
	Once again, the entanglement recovered is found to decrease as the interaction time increases where the generated spatial indistinguishability is not maximum. As in the amplitude damping scenario, such dephasing is monotonic in the Markovian regime and periodic in the non-Markovian one, with a decay rate which decreases as particle indistinguishability increases. Nonetheless, differently from that case, the entanglement now does not vanish. Indeed, for $t\to\infty$ it reaches a constant value which, under the above assumptions, is given by
	\begin{equation}
	C_\infty
	=\frac{2ll'rr'}{(lr')^2+(l'r)^2}.
	\end{equation}
	Furthermore, when the indistinguishability is maximum ($\mathcal{I}=1$) quantum correlations after the sLOCC measurement result to be completely immune to the action of the noisy environment and maintain their initial value.		
	Is is important to highlight that the existence of such a steady value for the entanglement of identical particles is only due to the spatial indistinguishability of the qubits and to the procedure used to produce the entangled state, i.e. the sLOCC operation. This result clearly shows that spatial indistinguishability of identical qubits can be exploited to recover quantum correlations spoiled by the detrimental noise of a phase damping-like environment interacting independently with the constituents, as shown in Figure\tilde\ref{phase_damping_diff}.
	
		\begin{figure}[t!]
		\includegraphics[width=0.48\textwidth]{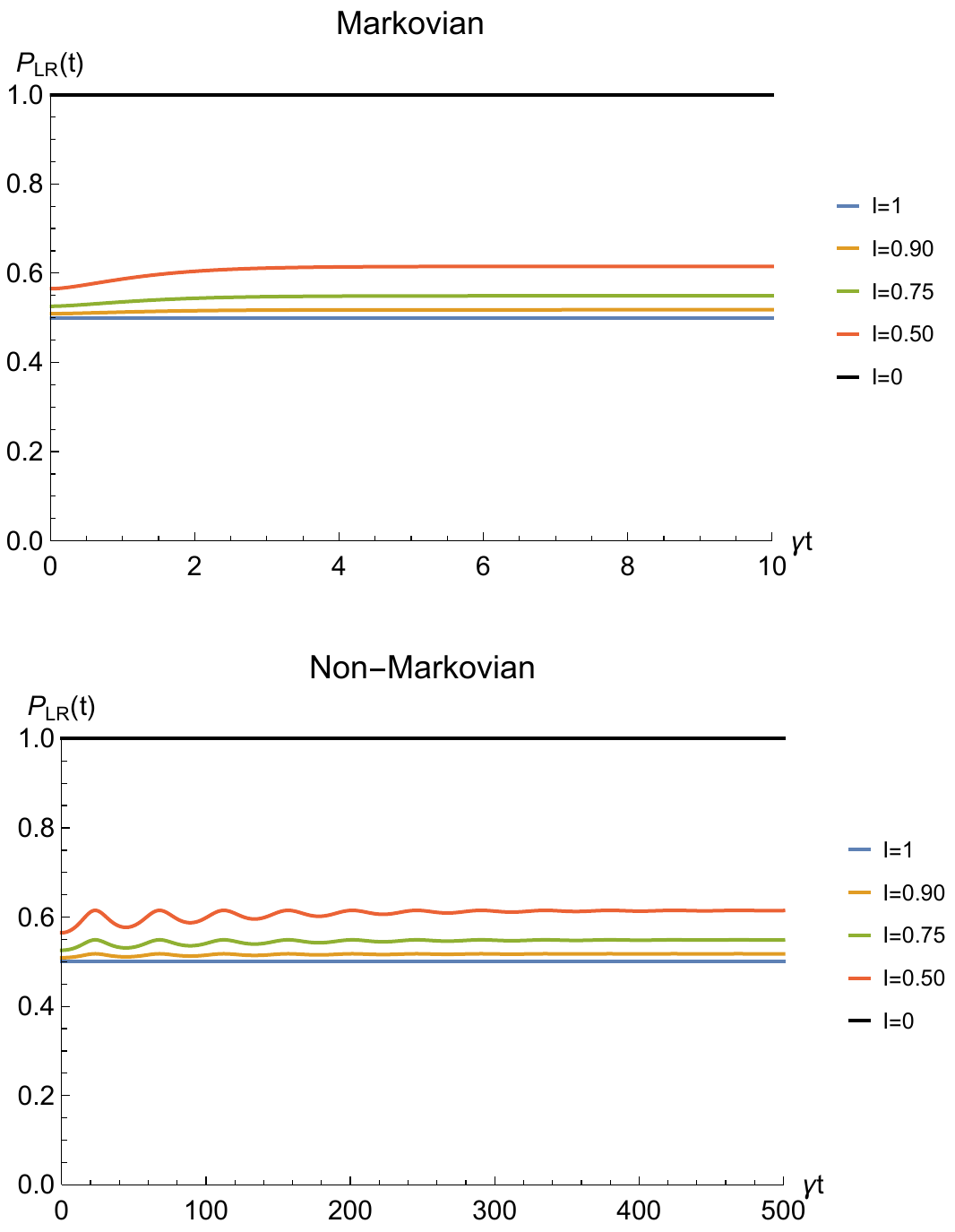}
		\caption{Probability of obtaining a non-zero outcome from the sLOCC projection for fermions  (with $l,l',r,r'>0$ and $l=r'$) interacting with localized phase damping channels. Different degrees of spatial indistinguishability are reported in both the Markovian ($\lambda=5\gamma$) (upper panel) and non-Markovian ($\lambda=0.01\gamma$) (lower panel) regimes.}
		\label{sloccprobpd}
	\end{figure}
	
	Finally, the success (sLOCC) probability of obtaining the outcome $\rlr(t)$ for two identical qubits undergoing local phase damping channels is
	\begin{equation}
	P_\text{LR}(t)
	=\frac{(lr')^2+(l'r)^2-2\eta\,ll'rr'\Big(1-p(t)\Big)}
	{\Big(1-\frac{1}{2}p(t)\Big)C_1^2+\frac{1}{2}p(t)C_2^2}.
	\end{equation}
	Figure\tilde\ref{sloccprobpd} depicts the behavior of the sLOCC probability of success\tilde\eqref{sloccprob} for fermions (with real and positive coefficients of the spatial wave functions) for different values of $\mathcal{I}$. 
	Once again, there is a trade-off between the probability of success and the concurrence, with $P_\text{LR}(t)=1$ when the particles are distinguishable and $P_\text{LR}=1/2$ for perfectly indistinguishable qubits. A similar general behavior is found for bosons (with the constraint $l=r'=l'=-r$), having $P_\text{LR}(t)=1-p(t)/2$ in the case of maximal indistinguishability $\mathcal{I}=1$ .

	\subsection{Depolarizing Channel}
	
	\begin{figure}[t!]
		\includegraphics[width=0.48\textwidth]{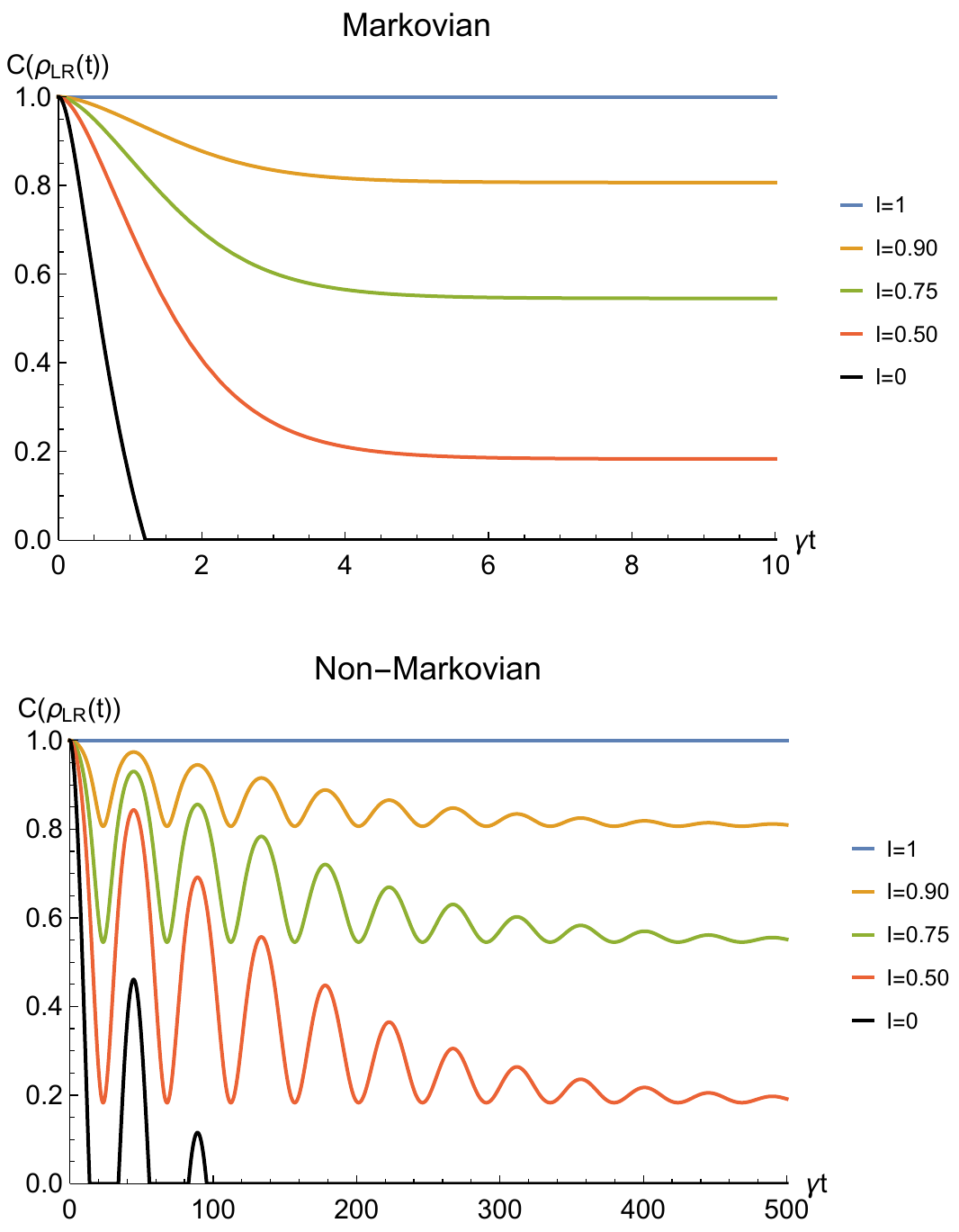}
		\caption{Concurrence of two identical qubits (fermions with $l,l',r,r'>0$, bosons with one of these four coefficients negative) in the initial state $\kom_\text{AB}$ subjected to localized depolarizing channels, undergoing an instantaneous deformation+sLOCC operation at time $t$ for different degrees of spatial indistinguishability $\mathcal{I}$ (with $|l|=|r'|$). Both Markovian ($\lambda=5\gamma$) (upper panel) and non-Markovian ($\lambda=0.01\gamma$) (lower panel) regimes are reported.}
		\label{depolarizing}
	\end{figure}

	In this section we reconsider and expand the results on entanglement protection at the preparation stage presented in Ref.~\cite{indistentanglprotection}. A depolarizing channel acting on a system of two qubits has the effect of leaving it untouched with probability $1-p(t)$ and of introducing a white noise which drives it into the maximally mixed state with probability $p(t)$. This is, for instance, a typical noise occurring when quantum states are initialized. 
	Supposing once again that our system of two identical particles is initially in the Bell state $\kom_\text{AB}$, it is well known that this kind of noisy interaction produces the Werner state \cite{nielsen2010quantum}
	\begin{equation}
	\label{noisystate3}
	\rab(t)
	=W^-_\text{AB}(t):
	=\Big(1-p(t)\Big)\kom_\text{LR}\bom_\text{LR}
	+\frac{1}{4}\,p(t)\,\id,
	\end{equation}
	where $1\!\!1$ is the $4\times4$ identity operator.
	Hereafter, we work for convenience on the Bell states basis
	\[
	\mathcal{B}_B
	=\{\kop_\text{AB},\kom_\text{AB},\ktp_\text{AB},\ktm_\text{AB}\},
	\]
	where $\kop_\text{AB},\,\kom_\text{AB}$ have been previously defined respectively in\tilde\eqref{bellsinglet} and\tilde\eqref{belltriplet}, while $\ktp_\text{AB}$ and $\ktm_\text{AB}$ are given by
	\begin{equation}
	\label{belltwo}
	\begin{gathered}
	\ktp_{\text{AB}}
	=\frac{1}{\sqrt{2}}\Big(\kaubu+\kadbd\Big),\\
	\ktm_{\text{AB}}
	=\frac{1}{\sqrt{2}}\Big(\kaubu-\kadbd\Big).
	\end{gathered}
	\end{equation}
	We recall that since such basis is orthonormal, the identity operator can be written as
	\[
	\id=\sum_{\substack{j=1,2\\s=\uparrow,\downarrow}}
	\ket{j_s}_\text{AB}\bra{j_s}_\text{AB}.
	\]
	
	At time $t$ we deform the two qubits wave functions. The deformation of states $\kop_\text{AB}$ and $\kom_\text{AB}$ has already been discussed in\tilde\eqref{deformed1+} and\tilde\eqref{deformed1-}, while states $\ktp_\text{AB}$ and $\ktm_\text{AB}$ get mapped respectively to
	\begin{equation}
	\label{deformed2}
	\dktp=C_2\nktp,
	\quad
	\dktm=C_2\nktm,
	\end{equation}
	where $\braket{\bar{2}_+|\bar{2}_+}_N=\braket{\bar{2}_-|\bar{2}_-}_N=1$ and $C_2$ is defined in\tilde\eqref{coeff2}.
	The result of the deformation of state\tilde\eqref{noisystate3} is thus the deformed Werner state of two indistinguishable qubits $\rho_D(t)
	=\bar{W}_D^-(t)$ \cite{indistentanglprotection}, where
	\begin{widetext}
	\begin{equation}
	\label{deformedstate3}
	\begin{aligned}
	\bar{W}_D^-(t)&
	:=\bigg[\Big(1-\frac{3}{4}\,p(t)\Big)\,C_1^2\,\nkom\nbom
	+C_2^2\,\,\frac{1}{4}\,p(t)\,\Big(\nkop\nbop+\nktp\nbtp+\nktm\nbtm\Big)\bigg]
	/
	\left[1-\eta\,\lvert\braket{\psi_1|\psi_2}\rvert^2\Big(1-\frac{3}{2}\,p(t)\Big)\right].
	\end{aligned}
	\end{equation}
	\end{widetext}
	To perform the final sLOCC measurement we assume that $\ket{\psi_1},\,\ket{\psi_2}$ have the usual structure given in Eq.\tilde\eqref{wfstructure}. Applying the projection operator on the state\tilde\eqref{deformedstate3} as defined in Eq.\tilde\eqref{sloccstate} we get
	\begin{widetext}
	\begin{equation}
	\label{rhoslocc3}
	\begin{aligned}
	\rlr(t)
	=&\ \bigg[\Big(1-\frac{3}{4}\,p(t)\Big)\,\absm\kom_\text{LR}\bom_\text{LR}
	+\frac{1}{4}\,p(t)\,\absp\Big(\kop_\text{LR}\bop_\text{LR}+\ktp_\text{LR}\btp_\text{LR}+\ktm_\text{LR}\btm_\text{LR}\Big)\bigg]\\&
	 /
	\left[\Big(1-\frac{3}{4}\,p(t)\Big)\absm+\frac{3}{4}\,p(t)\,\absp\right].
	\end{aligned}
	\end{equation}
	\end{widetext}
	Before computing the concurrence we notice that, as for the phase damping channel, the state of 
	Eq.\tilde\eqref{rhoslocc3} is real and diagonal on the Bell states basis, thus being invariant under the localized action of the Pauli matrices  $\sigma_y^L\otimes\sigma_y^R$. Therefore, the concurrence is evaluated in terms of the four eigenvalues of $\rlr(t)$, namely
	\[
	\begin{gathered}
	\lambda_A(t)
	=\frac{\Big(1-\frac{3}{4}\,p(t)\Big)\absm}
	{\Big(1-\frac{3}{4}\,p(t)\Big)\absm
		+\frac{3}{4}\,p(t)\,\absp},\\
	\lambda_j(t)=\frac{\frac{1}{4}\,p(t)\,\absp}
	{\Big(1-\frac{3}{4}\,p(t)\Big)\absm
		+\frac{3}{4}\,p(t)\,\absp},
	\end{gathered}
	\]
	where the index $j = B,C,D$.
	Considering once again fermions with real and positive coefficients $l,r,l',r'$, the concurrence has the expression
	\begin{equation}
	\label{concurrencedepreal}
	C\big(\rlr(t)\big)
	=\max\left\{
	0,\,
	\frac{\Big(1-\frac{3}{2}\,p(t)\Big)\Big[(lr')^2+(l'r)^2\Big]+2ll'rr'}
	{(lr')^2+(l'r)^2+\Big(1-\frac{3}{2}\,p(t)\Big)2ll'rr'}
	\right\}.
	\end{equation}

	\begin{figure}[t!]
		\includegraphics[width=0.48\textwidth]{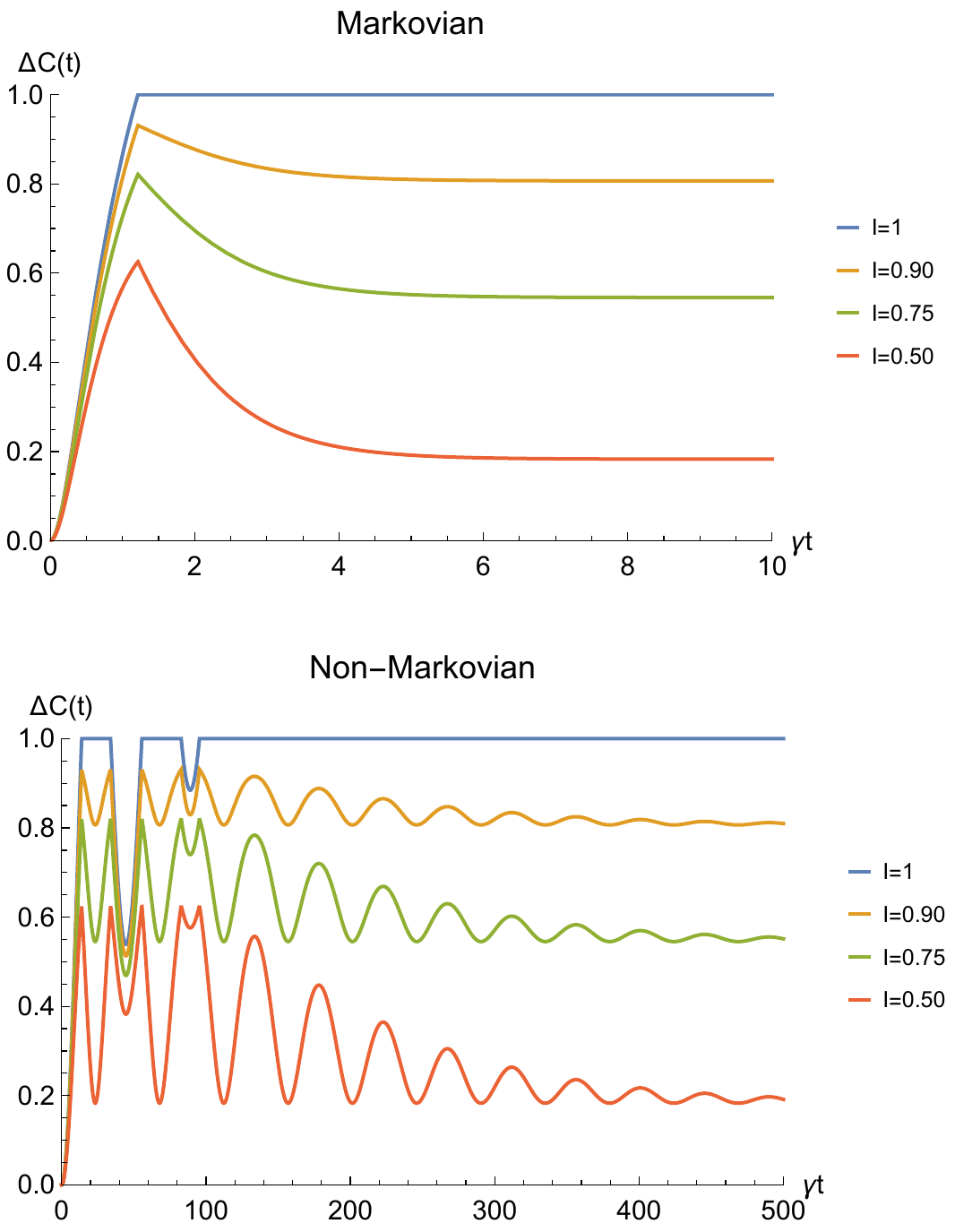}
		\caption{Net gain in the entanglement recovery of two identical qubits (fermions with $l,l',r,r'>0$, bosons with one of these four coefficients negative) in the initial state $\kom_\text{AB}$ interacting with a depolarizing channel, thanks to the deformation+sLOCC operation performed at time $t$. Results are reported for different degrees of spatial indistinguishability $\mathcal{I}$ (with $|l|=|r'|$). Both the Markovian ($\lambda=5\gamma$) (upper panel) and non-Markovian ($\lambda=0.01\gamma$) (lower panel) regimes are shown.}
		\label{depolarizing_diff}
	\end{figure}
	
Figure\tilde\ref{depolarizing} shows the time behavior of entanglement quantified by Eq.\tilde\eqref{concurrencedepreal}, while Figure\tilde\ref{depolarizing_diff} depicts $\Delta C(t)$. First of all, we emphasize that, differently from the amplitude damping channel and the phase damping channel, a sudden death phenomenon occurs when no deformation and sLOCC are performed: indeed, when $\mathcal{I}=0$ the entanglement vanishes at the finite time $\widetilde{t}$ such that $p(\widetilde{t})=2/3$. However, when $0<\mathcal{I}<1$, the state emerging from the sLOCC procedure recovers an amount of entanglement which decreases monotonically with $t$ in the Markovian regime and periodically in the non-Markovian regime. Nonetheless, as in the phase damping case, such decrease approaches a constant value given by
	\begin{equation}
	C_\infty
	=\max\left\{0,\,
	-\frac{(lr')^2+(l'r)^2-4\,ll'rr'}
	{2\big[(lr')^2+(l'r)^2-ll'rr'\big]}\right\}.
	\end{equation}
	Furthermore, we notice once again that when the maximum spatial indistinguishability ($\mathcal{I}=1$) is achieved, our procedure allows for a complete entanglement recovery independently on $t$.
	
	\begin{figure}[t!]
		\includegraphics[width=0.48\textwidth]{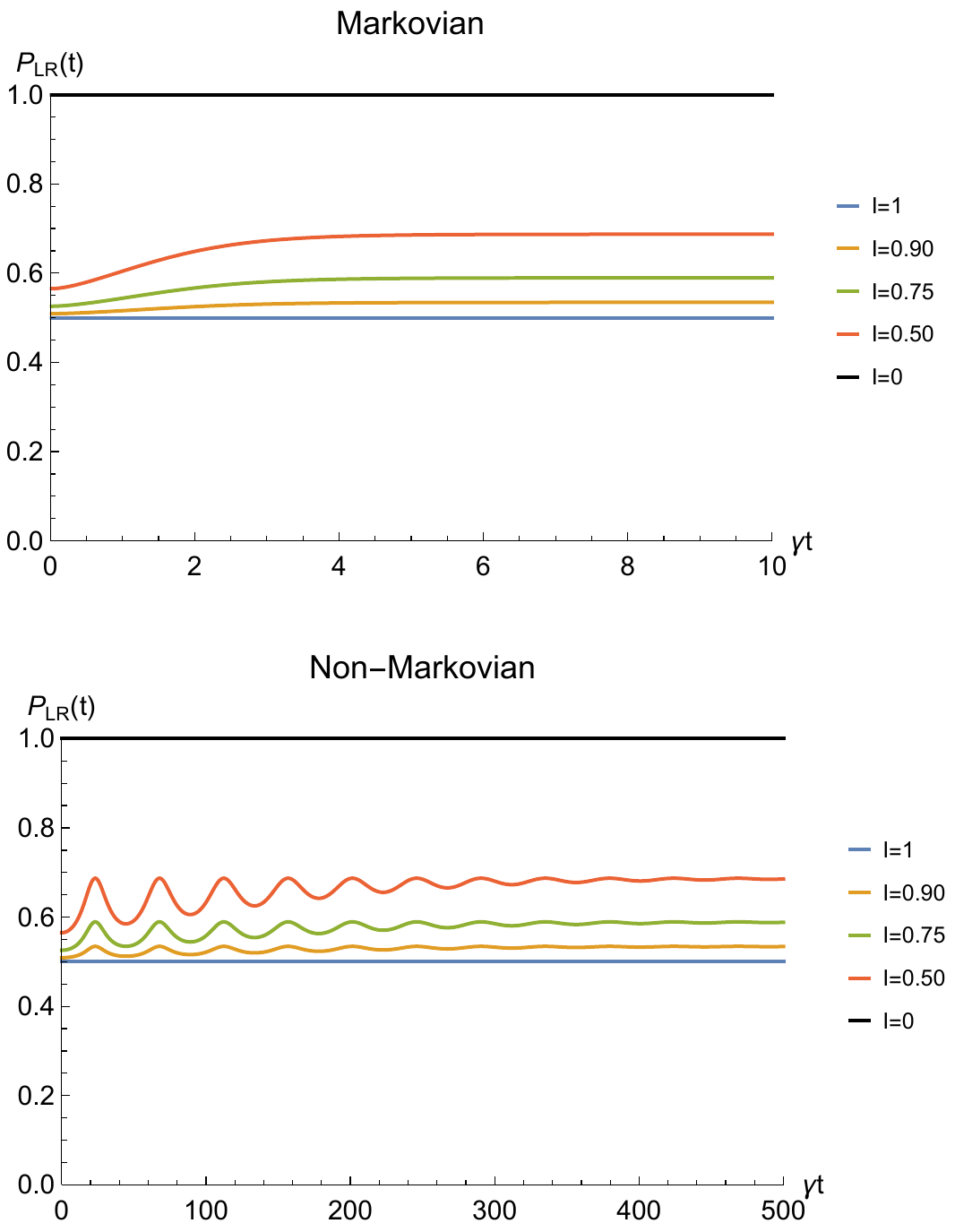}
		\caption{Probability of obtaining a nonzero outcome from the sLOCC projection for fermions with real and positive coefficients ($l=r'$) under a depolarizing channel. Different degrees of spatial indistinguishability $\mathcal{I}$ are reported in both Markovian ($\lambda=5\gamma$) (upper panel) and non-Markovian ($\lambda=0.01\gamma$) (lower panel) regimes.}
		\label{sloccprobdep}
	\end{figure}
	
	As a final quantity of interest we obtain the sLOCC probability of success, defined in Eq.\tilde\eqref{sloccprob}, for two identical qubits whose correlations have been spoiled by a local depolarizing channel, that is
	\begin{equation}
        P_\text{LR}(t)
    	=\frac{(lr')^2+(l'r)^2-2\eta\,ll'rr'\Big(1-\frac{3}{2}\,p(t)\Big)}
    	{1-\eta\Big[(ll')^2+(rr')^2+2ll'rr'\Big]\Big(1-\frac{3}{2}\,p(t)\Big)}.
    \end{equation}
	In Figure\tilde\eqref{sloccprobdep}, $P_\text{LR}(t)$ is plotted in the case of two fermions (with real and positive coefficients and $l=r'$) for different degrees of spatial indistinguishability. 
	Again, as expected, a trade-off exists between the probability of success and the concurrence, with the higher probability achieved when the qubits are perfectly distinguishable. Nonetheless, as happens in the previous channels, such probability reaches a stationary value which decreases as the indistinguishability increases, with $P_\text{LR}=1/2$ as the minimum value when $\mathcal{I}=1$. For bosons, a similar behavior is found (with the constraint $l=r'=l'=-r$), having $P_\text{LR}(t)=1-3p(t)/4$ when $\mathcal{I}=1$ \cite{indistentanglprotection}.

%%%%%%%%%%%%%%%%%%%%%%%%%%%%%%%%%%%%%%%%%%
\section{Discussion}\label{sec:Discussion}

In this paper we have shown that spatially localized operations and classical communication (sLOCC) provide an operational framework to successfully recover the quantum correlations between two identical qubits spoiled by the independent interaction with two noisy environments. The performance of such procedure is found to be strictly dependent on the degree of spatial indistinguishability reached by the spatial deformation of the particles wave functions. A general behavior has emerged: the higher is the degree of spatial indistinguishability, the better is the efficacy of the protocol, quantified by the difference between the amount of entanglement present at time $t$ with and without the application of our procedure. In particular, when the two particles are brought to perfectly overlap and the maximum degree of indistinguishability is achieved, the initial (maximum) amount of entanglement is completely recovered in all the considered scenarios, independently on how long the qubits have been interacting with the detrimental environment. 

If the indistinguishability is not maximum, instead, our results show that for an amplitude damping channel-like environment the entanglement after the sLOCC drops to zero after a short interaction time; nonetheless, the interval of time where the amount of recovered entanglement is significant increases with the indistinguishability in both the Markovian and the non-Markovian regimes. When the environment acts as a phase damping channel, instead, the recovered correlations are always nonzero and our protocol provides an exploitable resource independently on the interaction time (stationary entanglement). This behavior also holds in the depolarizing channel scenario, where the \emph{deformation+sLOCC protocol} achieves a special usefulness since it allows to recover quantum correlations destroyed at finite time by a sudden death phenomena.
	
We point out that the results reported in Figs.\tilde\ref{amplitude_damping},\tilde\ref{phase_damping},\tilde\ref{depolarizing} show a similar behavior to the ones discussed in 
Ref.~\cite{indistdynamicalprotection} (for a Markovian regime) where, in contrast to the present analysis, the system-environment interaction occurs between the deformation bringing the particles to spatially overlap and the final sLOCC measurements. Nonetheless, the decay rate is much larger in the situation considered here: the sLOCC operational framework for entanglement recovery performs better when the environment is not able to distinguish the particle it is interacting with, as happens in Ref.~\cite{indistdynamicalprotection}. Despite this, in a real world application it is likely that the system-environment interaction will occur both before the (spatial) deformation and between the deformation and the sLOCC. Therefore, an interesting possible prospect of this work would be to investigate the general open quantum system framework provided in Ref.~\cite{indistdynamicalprotection} when applied to noisy initial states such as those given in Eqs.\tilde\eqref{noisystate},\tilde\eqref{noisystate2}, and\tilde\eqref{noisystate3}. 

Our findings ultimately provide further insights about protection techniques of entangled states from the detrimental effects of surrounding environments by suitably manipulating the inherent indistinguishability of identical particle systems.

%\bibliography{Bibliografia}


\end{document}